\documentclass[%
 reprint,
 amsmath,amssymb,
 aps,
]{revtex4-1}

\usepackage{graphicx}
\usepackage{dcolumn}
\usepackage{bm}

\usepackage{color}

\begin{document}

\title{Phase-field modeling of isothermal quasi-incompressible multicomponent liquids}

\author{Gyula I. T\'oth}
\affiliation{Department of Physics and Technology, University of Bergen, All\'egaten 55, N-5007 Bergen, Norway}
\affiliation{Wigner Research Centre for Physics, P.O. Box 49, H-1525 Budapest, Hungary}
\email{Gyula.Toth@uib.no}


\date{\today}

\begin{abstract}
In this paper general dynamic equations describing the time evolution of  isothermal quasi-incompressible multicomponent liquids are derived in the framework of the classical Ginzburg-Landau theory of first order phase transformations. Based on the fundamental {equations of continuum mechanics}, a general convection-diffusion dynamics is set up first for compressible liquids. The constitutive relations for the diffusion fluxes and the capillary stress are determined in the framework of gradient theories. {Next the general definition of incompressibility is given}, which is taken into account {in the derivation} by using the Lagrange multiplier method. To validate the theory, the dynamic equations are solved numerically for the quaternary quasi-incompressible Cahn-Hilliard system. It is demonstrated that variable density (i) has no effect on equilibrium (in case of a suitably constructed free energy functional), {and (ii) can} influence non-equilibrium pattern formation significantly.   

\end{abstract}
   
\maketitle


\section{Introduction}

Although the theory of multicomponent liquid diffusion is well developed and {in use} in chemical engineering and materials science \cite{Cussler,TaylorKrishna,WesselinghKrishna,Datta20105976}, {the generalization of theory} describing the time evolution of {liquid systems} being out of mechanical equilibrium is still lacking. \textcolor{black}{Nevertheless, even results of simple continuum theories indicate that the hydrodynamic mode might influence pattern formation significantly even on small length scales \cite{Tegze2005418,Tegze20121689,C5CP02357B,PhysRevE.93.013126}, which indeed comes into play in microfluidics \cite{doi:10.1021/ma500302k,ANGE:ANGE201406040,nanoshell,PhysRevLett.108.198302},} {or even in oil industry related research: CO$_2$/water/hydrocarbon emulsions may offer a novel and ultimate solution for the global CO$_2$ storage problem with combined Enhanced Oil Recovery, however, the continuum description of kinetic processes in these molecularly complex liquids is nontrivial \cite{C5CP02357B}. Besides, significant volume changes occur in case of formation/dissociation of CO$_2$/CH$_4$ hydrates (clathrates), also necessitating thus a hydrodynamic description}. {Efforts regarding the effect of fluid flow on microscopic pattern formation phenomena} were made in the framework of diffuse interface models  {addressing various pattern formation problems including  fluid flow-assisted solidification, as well as binary and ternary liquid phase separation} \cite{Wheeler08081997,doi:10.1146/annurev.fluid.30.1.139,Lowengrub2617,Anderson2000175,PhysRevE.64.051601,PhysRevE.67.026117,PhysRevE.69.022601,KimLowengrub2005}. . {During pattern formation} in systems of practical importance interfaces move typically much slower than the speed of sound, {therefore}, the simplest (and most efficient) approaches include incompressible flow \cite{KimTernary,PhysRevE.93.013126}, {assuming} thus constant density. {In contrast, other theories} operate with {artificial} compressibilities ({resulting thus in a significant reduction in the speed of sound}) \cite{PhysRevE.64.051601,PhysRevE.67.026117,PhysRevE.69.022601} in order to access the time scale of solidification. The "golden mean" between the two {methods} can be the quasi-incompressible approach \cite{Anderson2000175,Lowengrub2617,KimLowengrub2005}, {in which} sound waves are {eliminated}, but density gap may exist between the bulk components. The problem of {this approach}, however, is the mathematical complexity {of the related numerical methods}, making {thus the theory unpractical} \cite{Guo2014486}. For these reasons, it is desirable to develop a physically consistent {and numerically efficient}, general dynamic framework describing the time evolution of multicomponent liquids. The essence of a general framework is keeping the coherency between the fundamental equations of continuum mechanics (the conservation of momentum and mass) and the principles of irreversible thermodynamics (the first and second law of thermodynamics, the Curie postulate, and the Onsager-Casimir theory). Besides the theoretical aspects, the model have practical importance only if it can handle variable density {on} pattern formation time scales, and can be solved {numerically efficiently}.\\
{The structure of the paper is as follows. In Section II,} general dynamic equations will be derived {on the basis of simple physical principles} for compressible liquids of arbitrary number of components. {In} Section III, a mathematically exact definition of general incompressibility is given.  All derivations are done in the framework of the classical Ginzburg-Landau theory of first order phase transformations. In Section IV, we apply the model for quasi-incompressible liquids, and derive an operator-splitting based, {pseudo}-spectral semi-implicit  {numerical} scheme to solve the dynamic equations. The incompressibility condition is handled by {a generalization of} Chorin's projection method. We show that a suitable discretization of the Navier-Stokes equation results in a generalized Poisson equation, which can be solved directly by using a two-step {pseudo}-spectral method. In Section V, a recently published multiphase-field model is used to address liquid-liquid phase separation in a quasi-incompressible multicomponent Cahn-Hilliard liquid, where the densities of the components, {as well as} the pairwise equilibrium interfacial properties can be calibrated independently from each other.  To demonstrate the physical consistency of the dynamic framework, we address equilibrium trijunctions and pattern formation in an asymmetric 4-component liquid in case of both constant and variable density, which is followed by the concluding remarks in Section VI.

\section{Compressible systems}

\subsection{Fundamental equations}

In continuum mechanics a mixture of $N$ liquids is characterized by the local and temporal \textit{mass} density [$\rho_i(\mathbf{r},t)$] and velocity [$\mathbf{v}_i(\mathbf{r},t)$] fields of the components ($i=1\dots N$). The conservation of mass and momentum read \cite{:/content/aapt/journal/ajp/16/2/10.1119/1.1991051,williams1958,:/content/aip/journal/pof2/18/7/10.1063/1.2221312}:
\begin{eqnarray}
\label{eq:massbase}&&\frac{\partial\rho_i}{\partial t}+\nabla\cdot(\rho_i\mathbf{v}_i)= 0 \enskip ;\\
\label{eq:mombase}&&\frac{\partial(\rho_i\mathbf{v}_i)}{\partial t}+\nabla\cdot(\rho_i\mathbf{v}_i\otimes\mathbf{v}_i)=\nabla\cdot\mathbb{S}_i+\mathbf{F}_i \enskip ,
\end{eqnarray}
where $\otimes$ stands for the tensor product. In Eq. (\ref{eq:mombase}),  the stress tensor $\mathbb{S}_i$ {accounts for self-specie} interactions, while $\mathbf{F}_i=\sum_{j\neq i}\mathbf{f}_{ij}$ is the net {"external"} force density {on component $i$} emerging solely from interspecie momentum exchanges (denoted by $\mathbf{f}_{ij}$). The time evolution of the system is driven by the stresses $\mathbb{S}_i$ and the force densities $\mathbf{F}_i$. Since these are abstract quantities {that} cannot be observed directly in experiments, a more convenient form of the dynamic equations is chosen. {Since Newton's third law prescribes} $\mathbf{f}_{ij}+ \mathbf{f}_{ji}=0$, $\mathbf{F}_i$ cancels in the sum of the momentum equations (i.e. the Navier-Stokes equation):
\begin{equation}
\label{eq:navierstokes}
\frac{\partial (\rho\,\mathbf{v})}{\partial t}+\nabla\cdot(\rho\,\mathbf{v}\otimes\mathbf{v}) = \nabla\cdot\mathbb{S} \enskip ,
\end{equation}
where $\rho(\mathbf{r},t)=\sum_i\rho_i(\mathbf{r},t)$ is the mixture density, and $\mathbf{v}=(1/\rho)\sum_i \rho_i\mathbf{v}_i$ is the velocity of the \textit{local center of mass}. Furthermore, the capillary stress reads $\mathbb{S} = \sum_i \left\{ \mathbb{S}_i - \rho_i [(\mathbf{v}_i-\mathbf{v}) \otimes (\mathbf{v}_i-\mathbf{v})] \right\}$. {Analogously to Eq. (\ref{eq:navierstokes}), summing up} Eq. (\ref{eq:massbase}) for all components results in the continuity equation:
\begin{equation}
\label{eq:continuity}
\frac{\partial \rho}{\partial t} + \nabla\cdot(\rho\,\mathbf{v}) = 0 \enskip .
\end{equation}   
Introducing {now} the \textit{mass fraction}s (or phase-fields) $c_i(\mathbf{r},t)=\rho_i(\mathbf{r},t)/\rho(\mathbf{r},t)$, and using Eq. (\ref{eq:continuity}) in Equations (\ref{eq:massbase}) and (\ref{eq:navierstokes}) {yields general dynamic equations of convection-diffusion type}:
\begin{equation}
\label{eq:finaleqns}\rho \frac{dc_i}{dt} = \nabla\cdot \mathbf{J}_i \enskip ; \quad \quad \rho \frac{d\mathbf{v}}{dt} = \nabla\cdot\mathbb{S} \enskip , 
\end{equation}
where $d/dt=\partial/\partial t+\mathbf{v}\cdot\nabla$ stands for the material derivative, and $\mathbf{J}_i=-\rho_i(\mathbf{v}_i-\mathbf{v})$ is the negative mass flux of component $i$ in the barycentric coordinate system, trivially resulting in the condition
\begin{equation}
\label{eq:bcond}
\sum_{i=1}^N \mathbf{J}_i = 0 \enskip .
\end{equation}
The phenomena "convection" and "diffusion" are postulated {here} in a complementary sense: {While} convection is attributed to the motion of the local center of mass, diffusion is interpreted as a motion relative to it.
 
\subsection{Constitutive relations}

\subsubsection{Diffusion fluxes}

Equations (\ref{eq:continuity}) and (\ref{eq:finaleqns}) govern the system via $\mathbf{J}_i$ and $\mathbb{S}$, for which constitutive equations have to be set up. Due to the Curie postulate, the diffusion process should be decoupled from the momentum transport in the framework of linear transport. Accordingly, the isotherm entropy production rate (time derivative of the entropy density) attributed to diffusion reads \cite{NYAS:NYAS241,:/content/aip/journal/jcp/73/3/10.1063/1.440202,nyiri}:
\begin{equation}
\label{eq:entropy}
T\,\dot{s} = \sum_{i=1}^N \mathbf{J}_i \cdot \nabla \lambda_i  \equiv \sum_{i=1}^N \mathbf{Q}_i \cdot \nabla \mu_i \enskip ,
\end{equation} 
where $\lambda_i$ is the \textit{specific} chemical potential (chemical potential per unit mass), the conjugate variable to the mass flux $\mathbf{J}_i$, whereas $\mu_i=M_i\,\lambda_i$ and $\mathbf{Q}_i=\mathbf{J}_i/M_i$ are the chemical potential and the molar flux, respectively ($M_i$ is the molar mass). According to the linearity postulate, the molar fluxes are the linear combination of the chemical potential {gradients}:
\begin{equation}
\label{eq:diff0}
\mathbf{Q}_i := \sum_{j=1}^N \alpha_{ij} \nabla\mu_j \enskip ,
\end{equation} 
where {$\alpha_{ij}$} is a symmetric matrix \cite{NYAS:NYAS241}. Furthermore, the second law of thermodynamics prescribes $T \dot{s} \geq 0$, which, together with Equations (\ref{eq:entropy}) and (\ref{eq:diff0}) indicate a positive semi-definite transport matrix. Using Eq. (\ref{eq:diff0}), together with the variable transformation yield
\begin{equation}
\label{eq:diff1}
\mathbf{J}_i := \sum_{j=1}^N L_{ij} \nabla\lambda_j \enskip ,
\end{equation}
where $L_{ij}=M_i M_j \alpha_{ij}$ is also symmetric (and positive semi-definite). Furthermore, using Eq. (\ref{eq:bcond}) results in 
\begin{equation}
\label{eq:Lcond}
\sum_{i=1}^N L_{ij} = \sum_{j=1}^N L_{ij} = 0 \enskip .
\end{equation}

\subsubsection{Stress tensor}

The stress tensor in Eq. (\ref{eq:finaleqns}) {is usually} divided into two contributions: 
\begin{equation}
\label{eq:capillary}
\mathbb{S}=\mathbb{R}+\mathbb{D} \enskip ,
\end{equation} 
where $\mathbb{R}$ is the non-classical reversible stress \cite{Korteweg1901}:
\begin{equation}
\label{eq:revstress}
\mathbb{R} = -P\,\mathbb{I} + \mathbb{A} \enskip ,
\end{equation}
where $P$ is the (scalar) thermodynamic pressure, $\mathbb{I}$ the identity matrix, while the Korteweg stress $\mathbb{A}$ can be determined by using the condition of mechanical equilibrium (also {cited as} generalized Gibbs-Duhem relation / least action principle in statistical physics) \cite{Salmon}:
\begin{equation}
\label{eq:GD}
\nabla\cdot\mathbb{R} = -\sum_{i=1}^N n_i\nabla\mu_i \equiv -\sum_{i=1}^N \rho_i\nabla\lambda_i \enskip ,
\end{equation}
where $n_i=\rho_i/M_i$ is the molar density. {Besides the reversible stress, t}he {other contribution to} $\mathbb{S}$ is the viscous stress $\mathbb{D}$. Assuming a Newtonian fluid {it can be approximated as}:
\begin{equation}
\label{eq:dissipative}
\mathbb{D} := \eta(\mathbf{c})\{[(\nabla\otimes\mathbf{v})+(\nabla\otimes\mathbf{v})^T]-(\nabla\cdot\mathbf{v})\} \enskip ,
\end{equation}
where
\begin{equation}
\label{eq:visc}
\eta(\mathbf{c}) := \sum_{i=1}^N \eta_i c_i
\end{equation} 
is the local (composition dependent) dynamic viscosity of the material. The dissipative stress defined by Eq. (\ref{eq:dissipative}) naturally satisfies $\mathbb{D}:\nabla\mathbf{v}\geq 0$ (the condition of non-negative entropy production) for $\eta(\mathbf{c}) \geq 0$ \cite{KimLowengrub2005}.

\subsection{Ginzburg-Landau Theory}

The Helmholtz free energy of the system ($F$) is often given as a functional of the local densities of the components. Since the density is conserved, the relevant thermodynamic potential of the system is the grand potential, which reads:
\begin{equation}
\label{eq:grandpot}
\Omega = F - \sum_{i=1}^N \mu_i^0 \int dV n_i \equiv F - \sum_{i=1}^N \lambda_i^0 \int dV \rho_i  \enskip ,
\end{equation}     
where $\mu_i^0=const.$ is the equilibrium chemical potential of component $i$ (analogously, $\lambda_i^0$ is the corresponding specific chemical potential). {The equilibrium configurations of the system can be found by solving the Euler-Lagrange equations prescribing constant chemical potentials:}
\begin{equation}
\label{eq:EL}
\frac{\delta \Omega}{\delta \rho_i}=0 \quad \Rightarrow \quad \frac{\delta F}{\delta \rho_i} = \lambda_i^0 \enskip ,
\end{equation}
where $\delta /\delta \rho_i$ denotes the functional derivative with respect to $\rho_i$. A suitable non-equilibrium generalization of the specific chemical potential {then} reads \cite{Carter1997}:
\begin{equation}
\label{eq:chempot}
\lambda_i:=\frac{\delta F}{\delta \rho_i} \enskip .
\end{equation}
Furthermore, since $\Omega=-P\,V$ in equilibrium, the equilibrium scalar pressure can be expressed from Eq. (\ref{eq:grandpot}), yielding $-P=f-\sum_{i=1}^N \mu_i^0 n_i \equiv f-\sum_{i=1}^N \lambda_i^0 \rho_i$, where $F=\int dV f${, and $f$ is evaluated here for an equilibrium solution}. This immediately indicates the non-equilibrium generalization of the pressure \cite{Wheeler08081997}:
\begin{equation}
\label{eq:pressure}
-P := f - \sum_{i=1}^N \rho_i\lambda_i  \enskip ,
\end{equation}
{where $\lambda_i$ is defined by Eq. (\ref{eq:chempot}).}
Assuming that the free energy is of the Ginzburg-Landau {type} \cite{PhysRevB.8.3423} (i.e. $f[\vec{\rho},\nabla\vec{\rho}]$ is given in terms of the densities and their gradients), and using Eq. (\ref{eq:pressure}) in Eq. (\ref{eq:GD}) result in the Korteweg stress \cite{Wheeler08081997}:
\begin{equation}
\label{eq:korteweg}
\mathbb{A} = -\sum_{i=1}^N \nabla\rho_i \otimes \frac{\partial f}{\partial \nabla\rho_i}  \enskip . 
\end{equation}

\subsection{Mass fraction formalism}

Since the dynamic equations defined by Equations (\ref{eq:continuity}) and (\ref{eq:finaleqns}) are written up for the total {mass} density and the mass fractions, it {would be} practical to express the diffusion fluxes and the {stress tensor} in terms of these variables. Since the chain rule of derivation applies also for the functional derivative (see Appendix B in Ref. \cite{PhysRevB.92.184105}, for example), one can write:
\begin{equation}
\label{eq:funcder}
\frac{\delta F}{\delta \rho_i} = \left(\frac{\delta F}{\delta \rho}\right)\left(\frac{\partial\rho}{\partial \rho_i}\right) + \sum_{j=1}^N \left(\frac{\delta F}{\delta c_j}\right)\left( \frac{\partial c_j}{\partial \rho_i}\right) \enskip .
\end{equation}
Using that $\partial \rho/\partial \rho_i=1$ and $\partial c_j/\partial \rho_i = (\delta_{ij}-c_j)/\rho$ (where $\delta_{ij}$ is the Kronecker-delta function {giving $1$ for $i=j$ and $0$ otherwise}), the specific chemical potential reads:
\begin{equation}
\label{eq:transform}
\lambda_i = \frac{\delta F}{\delta \rho_i} = \frac{1}{\rho} \left( \frac{\delta F}{\delta c_i} \right) + \left( \frac{\delta F}{\delta \rho} -\frac{\sigma}{\rho} \right) \enskip ,
\end{equation}
where $\sigma=\sum_{j=1}^N c_j (\delta F/\delta c_j)$. Note, that the term in the second bracket on the righ hand side is {independent from $i$}. Using Eq. (\ref{eq:transform}) in Eq. (\ref{eq:diff1}), and taking Eq. (\ref{eq:Lcond}) {also} into account results in
\begin{equation}
\label{eq:finalc1}
\mathbf{J}_i = \sum_{i=1}^N L_{ij} \nabla\left( \frac{1}{\rho} \, \frac{\delta F}{\delta c_i}\right) \enskip .
\end{equation}
Furthermore, using Eq. (\ref{eq:transform}) in Eq. (\ref{eq:pressure}) {yields} the scalar pressure:
\begin{equation}
\label{eq:finalc2}
\quad -P = f - \rho \frac{\delta F}{\delta \rho} \enskip .
\end{equation}
The {last} step of the derivation is to express the Korteweg stress defined by Eq. (\ref{eq:capillary}) as a function of $\rho$ and the mass fractions, which can be done analogously to Eq. (\ref{eq:funcder}), yielding:
\begin{equation}
\label{eq:kortewegf}
\mathbb{A} = -\nabla\rho\otimes\frac{\partial f}{\partial\nabla\rho} - \sum_{i=1}^N \nabla c_i \otimes \frac{\partial f}{\partial \nabla c_i} \enskip .
\end{equation}
Summarizing, the dynamic equations [see Eq. (\ref{eq:finaleqns})] together with the constitutive relations described by Equations 
(\ref{eq:capillary}), (\ref{eq:revstress}), (\ref{eq:dissipative}), (\ref{eq:visc}), (\ref{eq:finalc1}), (\ref{eq:finalc2}) and (\ref{eq:kortewegf}) define a complete convection-diffusion type {dynamic framework} for a given free energy functional $F[\rho,\mathbf{c},\nabla\rho,\nabla\mathbf{c}]$.

\section{Incompressible systems}

The constitutive equations derived in the previous section describe compressible systems, however, {incompressible systems have much higher practical importance in applications}, where time scales much larger than the time scale of sound waves are addressed. The {exact} \textit{mathematical} definition of incompressibility is that the {local} density is an {explicit} function of the local composition:
\begin{equation}
\label{eq:incomp}
\rho(\mathbf{r},t) \equiv \varrho[\mathbf{c}(\mathbf{r},t)] \enskip ,
\end{equation}
{where $\varrho(.)$ is a known function. Since} $\rho$ is not a free variable, $F[\mathbf{c},\nabla\mathbf{c}]$ is a functional only of the mass fractions {(and their gradients)}. Note that Eq. (\ref{eq:incomp}) together with Eq. (\ref{eq:continuity}) over-determine the dynamic equations. The degeneration of the solution can be eliminated by applying the Lagrange multiplier method as follows. The continuity equation can be {trivially} re-written in the form: $d\rho/dt=-\rho(\nabla\cdot\mathbf{v})$, where the material derivative of the density can be expressed now as $d\rho/dt = \sum_{i=1}^N (\partial \varrho/\partial c_i)(d c_i/dt)$. Furthermore, the diffusion equations indicate $\rho(dc_i/dt)=\nabla\cdot\tilde{\mathbf{J}}_i$, {resulting in}:
\begin{equation}
\label{eq:divv}
\nabla\cdot\mathbf{v} = \sum_{i=1}^N s_i \,(\nabla\cdot\tilde{\mathbf{J}}_i) \enskip ,
\end{equation}
where $s_i = -(1/\varrho^2)(\partial \varrho/\partial c_i)$, while the diffusion fluxes $\tilde{\mathbf{J}}_i$ ($i=1\dots N$) are to be defined later. Note that Eq. (\ref{eq:divv}) is an alternative form of the continuity equation for {general} incompressible systems. {It} prescribes a condition between the compositions and the velocity field, which can be taken into account by including the \textit{local} condition $\sum_{i=1}^N \rho_i(\mathbf{r},t) = \varrho(\{\rho_i(\mathbf{r},t)\})$ in the derivation of the non-equilbirium specific chemical potential. Accordingly, equilibrium represents a \textit{conditional} extremum of $\Omega$, yielding:
\begin{equation}
\label{eq:condEL}
\frac{\delta F}{\delta \rho_i} - \theta(\mathbf{r})\left(1-\frac{\partial \varrho}{\partial\rho_i} \right) = \tilde{\lambda_i}^0 \enskip ,
\end{equation} 
where $\theta(\mathbf{r})$ is the Langrange multiplier {attributed} to the local condition, whereas 
\begin{equation}
\label{eq:derivative}
\frac{\partial\varrho}{\partial \rho_i}=\sum_{j=1}^N \frac{\partial\varrho}{\partial c_j}\frac{\partial c_j}{\partial \rho_i}=\frac{1}{\varrho}\left(\frac{\partial \varrho}{\partial c_i} - \sum_{j=1}^N \frac{\partial \varrho}{\partial c_j} c_j \right)
\end{equation} 
is non-trivial, and must be calculated from the known $\varrho(\mathbf{c})$ relationship. {Accordingly, t}he (conditional) specific chemical potential {reads}
\begin{equation}
\label{eq:condpot}
\tilde{\lambda}_i := \frac{\delta F}{\delta \rho_i} - \Theta(\mathbf{r},t)\left(1-\frac{\partial\varrho}{\partial\rho_i} \right) \enskip .
\end{equation}
Note, that the relevant thermodynamic force is {now} $\nabla\tilde{\lambda}_i$, since {it} becomes constant in equilibrium {(according to Eq. (\ref{eq:condEL})}. The diffusion fluxes are then postulated as:
\begin{equation}
\label{eq:condflux}
\tilde{\mathbf{J}}_i:=\sum_{j=1}^N L_{ij}\nabla\tilde{\lambda}_j \enskip .
\end{equation}
Analogously to the chemical potential, the conditional pressure reads  $-\tilde{P}:=f-\sum_{i=1}^N \rho_i\tilde{\lambda}_i$. {Expanding the expression yields}: 
\begin{equation}
\label{eq:condpres}
-\tilde{P} = -P + \Theta(\mathbf{r},t)\,\left(\varrho-\sum_{i=1}^N \rho_i \frac{\partial\varrho}{\partial\rho_i}\right) \enskip ,
\end{equation}
where $P$ is defined by Eq. (\ref{eq:pressure}). The conditional reversible stress then reads
\begin{equation}
\label{eq:condrev}
\tilde{\mathbb{R}}:=-\tilde{P}\,\mathbb{I}+\mathbb{A} \enskip ,
\end{equation}
where $\mathbb{A}$ is defined by Eq. (\ref{eq:korteweg}). Comparing Equations (\ref{eq:condpres}) and (\ref{eq:condrev}) results in  $\nabla\cdot\tilde{\mathbb{R}}=-\sum_{i=1}^N\rho_i\nabla\tilde{\lambda}_i$ (a conditional Gibbs-Duhem relationship), correctly accounting for mechanical equilibrium. {Furthermore, u}sing Equations (\ref{eq:transform}), (\ref{eq:finalc2}) and (\ref{eq:derivative}) in Equations (\ref{eq:condpot}) and (\ref{eq:condpres}) results in {the conditional specific chemical potential expressed in terms of the mass fractions}:
\begin{equation}
\label{eq:const1}\tilde{\lambda_i} = \frac{1}{\varrho}\frac{\delta F}{\delta c_i} + (f+\tilde{P}) s_i \enskip ,
\end{equation}
where all terms canceling in Eq. (\ref{eq:condflux}) {due to Eq. (\ref{eq:Lcond})} are neglected.  Note that the appearance of the pressure in Eq. (\ref{eq:const1}) expresses the fact that inertia must be taken into account in case of varying density \cite{personal}. Finally, the Korteweg stress simply reads:
\begin{equation}
\label{eq:kst}
\mathbb{A} = - \sum_{i=1}^N \nabla c_i \otimes \frac{\partial f}{\partial \nabla c_i} \enskip ,
\end{equation}
which follows from the fact that $F$ is not a function of $\rho$.

\section{Chorin's projection method for quasi-incompressible liquids}

{It is very practical to test the theory for} quasi-incompressible liquids. {In these systems,} the bulk components are incompressible, {while they can have} different densities. {In addition, Amagat's law of additive volumes also applies for these systems, i.e. the partial volumes of the components do not change upon mixing}. These conditions can be expressed in terms of the mass fractions as \cite{KimLowengrub2005}:
\begin{equation}
\label{eq:quasi}
\frac{1}{\varrho} = \sum_{i=1}^N \frac{c_i}{\rho_i^0} \enskip ,
\end{equation}
where $\rho_i^0$ is the {bulk} density of component $i$. From Eq. (\ref{eq:quasi}) it follows that 
\begin{equation}
s_i=-\frac{1}{\varrho^2}\frac{\partial \varrho}{\partial c_i} = \frac{1}{\rho_i^0}
\end{equation}
is constant in Equations (\ref{eq:divv}) and (\ref{eq:const1}). In solving the dynamic equations numerically, we face two major problems: (i) The mass fractions and the velocity are non-conserved and (ii) the pressure appears in both the conditional specific chemical potentials and the reversible stress tensor. (i) can be resolved by taking into account that
\begin{equation}
\frac{\partial (y\varrho)}{\partial t}+\nabla \cdot [(y\varrho)\mathbf{v}]\equiv \varrho \frac{dy}{dt}
\end{equation}
(where $y=c_i,\mathbf{v}$), simply because of the continuity equation [see Eq. (\ref{eq:continuity})]. Therefore, the dynamic equations can be re-written as:
\begin{eqnarray}
\label{eq:trans1}\frac{\partial \rho_i}{\partial t} &=& \nabla\cdot\left( \tilde{\mathbf{J}}_i - \varrho\,c_i\mathbf{v} \right) \\
\label{eq:trans2}\frac{\partial \mathbf{p}}{\partial t} &=& \nabla\cdot\left( \tilde{\mathbb{R}}+\mathbb{D} - \varrho\,\mathbf{v}\otimes\mathbf{v} \right) \enskip ,
\end{eqnarray}
where the right hand sides are still expressed in terms of the mass fractions and the velocity field, which are {explicitly} related to the absolute densities and the momentum via {$\mathbf{p}=\rho\,\mathbf{v}$ and} Eq. (\ref{eq:quasi}). Eq. (\ref{eq:quasi}) can {also} be expressed in terms of the absolute densities, yielding $1=\sum_{i=1}^N(\rho_i/\rho_i^0)$. Consequently, dividing Eq. (\ref{eq:trans1}) by $\rho_i^0$, then {summing them up} trivially results in Eq. (\ref{eq:divv}), an alternative form of continuity for prescribed total density. It means, that once $\sum_{i=1}^N (\rho_i/\rho_i^0)=1$ \textit{and} Eq. (\ref{eq:divv}) are satisfied, Eq. (\ref{eq:quasi}) always applies for Eq. (\ref{eq:trans1}). To solve the dynamic equations we apply an operator-splitting based, {pseudo}-spectral semi-implicit method, {yielding} \cite{Tegze20091612,PhysRevE.93.013126}:
\begin{eqnarray}
\label{eq:scheme1}
\rho_i^{t+\Delta t}(\mathbf{k}) &:=& \rho_i^{t}(\mathbf{k}) + \frac{\Delta t}{1+\Delta t\,S_c(k)}[(\imath\,\mathbf{k}) \cdot \mathbf{b}_i^t(\mathbf{k})]  \\
\label{eq:scheme2}\mathbf{p}^*(\mathbf{k}) &:=& \mathbf{p}^{t}(\mathbf{k}) + \frac{\Delta t}{1+\Delta t\,S_v(k)}[(\imath\,\mathbf{k}) \cdot \mathbb{B}^t(\mathbf{k})] \enskip ,
\end{eqnarray}
where $\mathbf{k}$ is the wave number, $\Delta t$ the time step, $\mathbf{b}_i^t(\mathbf{k})$ and $\mathbb{B}^t(\mathbf{k})$ the Fourier transform of the right hand sides behind the divergence {in} Equations (\ref{eq:trans1}) and (\ref{eq:trans2}), respectively, while $S_{c}(k)$ and $S_v(k)$ are suitably chosen splitting operators (for details, see Reference \cite{PhysRevE.93.013126}). Following Chorin's projection method \cite{Chorin}, the momentum {density given by Eq. (\ref{eq:scheme2})} should be corrected as:
\begin{equation}
\label{eq:scheme3}
\mathbf{p}^{t+\Delta t}(\mathbf{r}) :=\mathbf{p}^*(\mathbf{r}) - \Delta t\, [ \nabla \delta P(\mathbf{r})] \enskip ,
\end{equation}
where $P^{t+\Delta t}(\mathbf{r}):=P^t(\mathbf{r})+\delta P(\mathbf{r})$ is the new pressure. Note that Equations (\ref{eq:scheme1})-(\ref{eq:scheme3}) ensure the conservation of the {mass and the momentum densities} numerically \textit{exactly}. The pressure correction $\delta P$ can be determined from Eq. (\ref{eq:divv}) and Eq. (\ref{eq:scheme3}) by taking into account that $\mathbf{p}^{t+\Delta t}\equiv \varrho^{t+\Delta t}\,\mathbf{v}^{t+\Delta t}$, yielding
\begin{equation}
\nabla\cdot\mathbf{v}^{t+\Delta t} = \nabla \cdot \left( \mathbf{v}^* - \frac{\Delta t\,\nabla\delta P}{\varrho^{t+\Delta t}} \right) = \sum_{i=1}^N \frac{\nabla\cdot\tilde{\mathbf{J}}_i^{t+\Delta t}}{\rho_i^0}
 \enskip ,
\end{equation}
where $\mathbf{v}^* \equiv \mathbf{p}^*/\varrho^{t+\Delta t}$. Note, that $\tilde{\mathbf{J}}_i^{t+\Delta t}$ contains $P^{t+\Delta t}$ because of Eq. (\ref{eq:const1}). Introducing $\mathbf{J}^*_i$ as $\tilde{\mathbf{J}}_i$ evaluated for $\mathbf{c}^{t+\Delta t}$ and $P^t$ results in
\begin{equation}
\label{eq:incomp2}
\nabla \cdot \mathbf{v}^* - \sum_{i=1}^N \frac{\nabla\cdot \mathbf{J}^*_i}{\rho_i^0} = \nabla \cdot \left[ \left( \frac{\Delta t}{\varrho^{t+\Delta t}} + \sum_{i,j}\frac{L_{ij}^{t+\Delta t}}{\rho_i^0\rho_j^0} \right) \nabla\delta P \right] \enskip ,
\end{equation}
where a general transport matrix $\mathbb{L}$ having composition dependent elements is assumed. Note that Eq. (\ref{eq:incomp}) is a $h(\mathbf{r}) = \nabla \cdot[g(\mathbf{r})\nabla f(\mathbf{r})]$ type equation, which, assuming periodic boundary conditions, can {also} be solved by {applying} a two-step {pseudo}-spectral method as follows. First, the outer equation $\nabla \cdot \mathbf{z}(\mathbf{r})=h(\mathbf{r})$ is solved, yielding:
\begin{equation}
\mathbf{z}(\mathbf{k}) = \frac{h(\mathbf{k})}{\imath\,k^2}\mathbf{k} \enskip .
\end{equation}
Having $\mathbf{z}(\mathbf{k})$, $\mathbf{z}(\mathbf{r})=g(\mathbf{r})\nabla f(\mathbf{r})$ {can be} solved for $f(\mathbf{k})$. Introducing $\mathbf{w}(\mathbf{r}):=\mathbf{z}(\mathbf{r})/g(\mathbf{r})$ results in:
\begin{equation}
f(\mathbf{k}) = \frac{\mathbf{k} \cdot \mathbf{w}(\mathbf{k})}{\imath\,k^2} \enskip .
\end{equation} 
In our problem, $h(\mathbf{r})$ is the left hand side of Eq. (\ref{eq:incomp}), $g(\mathbf{r})$ is the term in the round bracket on the right hand side, and $f(\mathbf{r})\equiv \delta P(\mathbf{r})$. Having $\delta P(\mathbf{k})$, the momentum update described by Eq. (\ref{eq:scheme3}), together with the pressure update $P^{t+\Delta t}(\mathbf{k})=P^t(\mathbf{k})+\delta P(\mathbf{k})$ can be directly applied. In practice, the time stepping scheme contains the following steps:
\begin{enumerate}
\item {calculating} $c_i^t(\mathbf{r})=\rho_i^t(\mathbf{r})/\varrho^t(\mathbf{r})$ and $\mathbf{v}^t(\mathbf{r})=\mathbf{p}^t(\mathbf{r})/\varrho^t(\mathbf{r})$ from $\rho_i^t(\mathbf{r})$ and $\mathbf{p}^t(\mathbf{r})$
[where $\varrho^t(\mathbf{r})=\sum_{k=1}^N \rho_k^t(\mathbf{r})$];
\item evaluation of {$\mathbf{b}_i^t(\mathbf{r})$ and $\mathbb{B}^t(\mathbf{r})$} appearing on the right hand sides of Equations (\ref{eq:scheme1}) and (\ref{eq:scheme2}), {respectively};
\item time stepping to generate $\rho_i^{t+\Delta t}(\mathbf{k})$ and $\mathbf{p}^*(\mathbf{k})$;
\item calculating the new variables $c_i^{t+\Delta t}(\mathbf{r})=\rho_i^{t+\Delta t}(\mathbf{r})/\varrho^{t+\Delta t}(\mathbf{r})$ and {the intermediate velocity field} $\mathbf{v}^*(\mathbf{r})=\mathbf{p}^*(\mathbf{r})/\varrho^{t+\Delta t}(\mathbf{r})$;
\item solving Eq. (\ref{eq:incomp2}) for $\delta P(\mathbf{k})$;
\item updating the pressure and the momentum {density} to have $P^{t+\Delta t}(\mathbf{k})$ and $\mathbf{p}^{t+\Delta t}(\mathbf{k})$.
\end{enumerate}
Finally we note, that solving Eq. (\ref{eq:incomp}) is "safe", since the term in the round bracket on the right hand side [i.e. $g(\mathbf{r})$] is strictly positive, due to the fact that $\mathbb{L}$ is a positive semi-definite matrix.

\section{Cahn-Hilliard liquids}

\subsection{Free energy, dynamic equations and scaling}

The free energy density of {a} general multicomponent liquid undergoing liquid-liquid phase separation can be defined as \cite{PhysRevB.92.184105,PhysRevE.93.013126}:
\begin{equation}
\label{eq:spec}
f(\mathbf{c},\nabla\mathbf{c}) := [w(\mathbf{c}) g(\mathbf{c}) + A_3 f_3(\mathbf{c})] + \frac{\epsilon^2(\mathbf{c})}{2} \sum_{i=1}^N (\nabla c_i)^2 \enskip ,
\end{equation}
where
\begin{eqnarray}
\nonumber &&g(\mathbf{c}) = \frac{1}{12}+\sum_{i=1}^N \left( \frac{c_i^4}{4}-\frac{c_i^3}{3} \right) + \frac{1}{2}\sum_{i<j} c_i^2c_j^2 \\
\label{eq:details}&&f_3(\mathbf{c}) = \sum_{i<j<k} |c_i||c_j||c_k| \\
\nonumber &&w(\mathbf{c}) = \frac{\sum_{i<j} w_{ij} (c_ic_j)^2}{\sum_{i<j}(c_i c_j)^2} \enskip ; \quad \epsilon^2(\mathbf{c}) = \frac{\sum_{i<j} \epsilon^2_{ij} (c_ic_j)^2}{\sum_{i<j}(c_i c_j)^2} \enskip ,
\end{eqnarray}
while the transport matrix reads:
\begin{eqnarray}
\label{eq:transport}
L_{ij} &=& - \frac{\kappa_{ij}}{2}\left| \frac{c_i}{1-c_i}\right|\left| \frac{c_j}{1-c_j}\right| \quad \text{for} \quad i\neq j \\
\nonumber L_{ii} &=& - \sum_{j \neq i} L_{ij} \enskip ,
\end{eqnarray}
with $\kappa_{ij}>0$ pairwise diffusion mobilities. The model parameters $w_{ij}$ and $\epsilon^2_{ij}$ can be related to the interfacial tension $\sigma_{ij}$ and the interface thickness $\delta_{ij}$ of the pure binary planar $(i,j)$ interface as:
\begin{equation}
w_{ij}=3\,\sigma_{ij}/\delta_{ij} \quad \text{and} \quad \epsilon^2_{ij}=3\,\sigma_{ij}\,\delta_{ij} \enskip .
\end{equation}
Furthermore, the pairwise mobilities $\kappa_{ij}$ can be related to the diffusion coefficents as follows. {Having} the diffusion coefficient $D_{ij}$ of component $i$ in bulk component $j$, a \textit{single} mobility $\kappa_{ij}$ might be defined as:
\begin{equation}
\label{eq:mobility}
\kappa_{ij} := \rho_i^0 \rho_j^0 \bar{D}_{ij}/w_{ij} \enskip , 
\end{equation}
where $\bar{D}_{ij}:=\sqrt{D_{ij}D_{ji}}$. {In} gradient theories the diffusion equations reduce to the Lorentzian limit {close to equilibrium}:
\begin{equation}
\label{eq:Lorentzian}
\frac{\partial \delta c_{i,j}}{\partial t} = \frac{\kappa_{ij}w_{ij}}{(\rho_{j,i}^0)^2} [\nabla^2 (1+\omega_{ij}\nabla^2) ] \delta c_{i,j} 
\end{equation}
for $\mathbf{v}\to 0$, $c_i(\mathbf{r},t)+c_j(\mathbf{r},t)=1$, $\delta c_{i,j}(\mathbf{r,t}) \to 0$ and $c_{k\neq i,j}(\mathbf{r},t)\equiv 0$. 
Using Eq. (\ref{eq:mobility}) in Eq. (\ref{eq:Lorentzian}), the modified diffusion coefficients read:
\begin{equation}
D'_{ij} = (\rho_i^0/\rho_j^0) \bar{D}_{ij} \quad \text{and} \quad D'_{ji} = (\rho_j^0/\rho_i^0) \bar{D}_{ij}
\end{equation}
(approximates of the original ones). In {a recently published work}  \cite{PhysRevE.93.013126} it has been shown that the construction described by Equations (\ref{eq:spec})-(\ref{eq:transport}) has the following advantages:
\begin{itemize}
\item the model naturally reduces / extends on the level of both the free energy functional and the dynamics when removing / adding a phase, respectively;
\item the pure planar binary interfaces $c_i(x)=\{1+\tanh[x/(2 \delta_{ij})]\}/2$, $c_{j \neq i}=1-c_i(x)$ and $c_{k\neq i,j}$ minimize the free energy functional with interfacial tension $\sigma_{ij}$;
\item the system shows the $f_{bulk} < f_{interface} < f_{trijunction} < \dots$ energy hierarchy, i.e. it {hierarchically penalize local multicomponent} states;
\item the appearance of spurious phases are excluded.
\end{itemize}
These features apply {in case of} \textit{constant} density. To address {variable density}, the structure of equilibrium solutions is investigated first. The general Euler-Lagrange equations follow from substituting Eq. (\ref{eq:transform}) into (\ref{eq:condEL}), and using Eq. (\ref{eq:derivative}){, yielding}:
\begin{equation}
\label{eq:quasiEL}
\nabla \left[ \frac{1}{\varrho}\frac{\delta F}{\delta c_i} + \frac{\lambda(\mathbf{r})}{\rho_i^0} \right] = \nabla \left[ \frac{1}{\varrho} \frac{\delta F}{\delta c_j} + \frac{\lambda(\mathbf{r})}{\rho_j^0} \right]
\end{equation}
for $i,j=1\dots N$, where $\lambda(\mathbf{r})$ is the equilibrium value of $f+\tilde{P}$. Since for all solutions of the Euler-Lagrange equations the functional derivatives $\delta F/\delta c_i$ vanish \cite{PhysRevE.93.013126}, the solutions of the original model represent solutions of Eq. (\ref{eq:quasiEL}) too for arbitrary $\lambda(\mathbf{r})=const$. Therefore, the equilibrium solutions {of the present model} {coincide} for constant and variable density.\\
  
Choosing now the length scale $\delta$, the time scale $\delta^2/D$ (where $D$ is the scale of the diffusion coefficients), the density scale $\rho$, and {the} interfacial tension scale $\sigma$, and the free energy density scale $3\,\sigma/\delta$ result in the dimensionless diffusion equations:
\begin{equation}
\label{eq:scaleddiff}
\hat{\varrho} \frac{d c_i}{d \hat{t}} = \hat{\nabla} \cdot \hat{\mathbf{J}}_i \enskip ,
\end{equation}
where
\begin{eqnarray}
\nonumber \hat{\mathbf{J}}_i &=& \sum_{i=1}^N \hat{L}_{ij} \hat{\nabla} \hat{\lambda}_i \enskip ; \quad \quad \enskip \hat{\lambda}_i = \frac{1}{\hat{\varrho}}\frac{\delta \hat{F}}{\delta c_i} + \frac{\hat{f}+\hat{P}}{\hat{\rho}_i^0} \\
\label{eq:scaledJ}\frac{\delta \hat{F}}{\delta c_i} &=& \frac{\partial \hat{f}}{\partial c_i} - \hat{\nabla} \frac{\partial \hat{f}}{\partial \hat{\nabla} c_i} \\
\nonumber \hat{f} &=& \hat{w}(\mathbf{c})g(\mathbf{c}) + \tilde{A}_3 f_3(\mathbf{c}) + \frac{\hat{\epsilon}^2(\mathbf{c})}{2} \sum_{i<j} (\hat{\nabla}c_i)^2 \enskip ,
\end{eqnarray}
where $\hat{w}_{ij}=\hat{\sigma}_{ij}/\hat{\delta}_{ij}$, $\hat{\epsilon}^2_{ij}=\hat{\sigma}_{ij}\hat{\delta}_{ij}$ [see Eq. (\ref{eq:details})], and $\hat{\kappa}_{ij}:=\hat{\rho}_i^0\hat{\rho}_j^0\hat{\bar{D}}_{ij}/\hat{w}_{ij}$  [see Eq. (\ref{eq:mobility})]. The dimensionless Navier-Stokes equation reads
\begin{equation}
\hat{\varrho}\frac{d\hat{\mathbf{v}}}{d\hat{t}} = \hat{\nabla} \cdot ( \beta_R\hat{\mathbb{R}}+\beta_D\hat{\mathbb{D}}) \enskip ,
\end{equation}
where 
\begin{equation}
\beta_R = \frac{3\,\sigma\,\delta}{D^2\rho} \quad \text{and} \quad \beta_D=\frac{\eta}{\rho\,D} \enskip ,
\end{equation}
and $\eta$ is the viscosity scale. The reversible and irreversible components of the dimensionless stress tensor read: 
\begin{eqnarray}
\hat{\mathbb{R}} &=& -\hat{P}\,\mathbb{I} - \hat{\epsilon}^2(\mathbf{c}) \sum_{i=1}^N (\hat{\nabla}c_i \otimes \hat{\nabla}c_i) \\
\hat{\mathbb{D}} &=& \hat{\eta}(\mathbf{c})[(\hat{\nabla}\otimes\mathbf{v})+ (\hat{\nabla}\otimes\mathbf{v})^T] - (\hat{\nabla}\cdot\hat{\mathbf{v}})\,\mathbb{I} \enskip .
\end{eqnarray}
Finally, the incompressibility condition simply transforms into
\begin{equation}
\label{eq:scaledcond}
\hat{\nabla}\cdot \hat{\mathbf{v}} = \sum_{i=1}^N \frac{\hat{\nabla}\cdot\hat{\mathbf{J}}_i}{\hat{\rho}_i^0} \enskip .
\end{equation}
It's straightforward to see the role of the dimensionless model parameters $\beta_R$ and $\beta_D$: The {dynamics} is diffusion dominated when $\beta_R/\beta_D \propto \frac{\sigma \delta }{D \eta}\ll 1$, since any flow generated by the {capillary} stress {is suppressed}. In contrast, for $\beta_R \gg 1$ together with $\beta_D/\beta_R \ll 1$, the {dynamics} is fluid flow dominated, since {convection} overwhelms diffusion.   

\begin{table}[h]
\caption{Characteristic model parameters for 2 system {types}.}
\begin{tabular}{|c|c|c|c|c|c|}
\hline
type & $\rho$ [kg/m$^3$] & $\sigma$ [mJ/m$^2$] & $\delta$ [\AA] & $D$ [$\mu$m$^2$/s] & $\eta$ [mPas] \\
\hline
\hline
I & 1000 & 50 & 1 & 500 & 1 \\
\hline
II & 5000 & 200 & 10 & 1000 & 0.5 \\
\hline
\end{tabular}
\end{table}

\subsection{Results of numerical simulations}

The numerical solution of the dimensionless equations have been done in 2 dimensions on $2048\times1024$ and $1024\times1024$ computational grids with grid spacing $\Delta \hat{x}=1/2$ and time step $\Delta \hat{t}=5\times10^{-3}$. The physical parameters {used} mimic a typical water/hydrocarbon/CO$_2$ system at room temperature and high pressure (system type I in Table I). For comparison, typical parameters for liquid metals at high temperature and atmospheric pressure were also considered (system type II in Table I). The resulting scale parameters prescribe fluid flow dominated {dynamics} for both types. In our numerical simulations, we chose $\beta_R=6 \times 10^{4}$ and $\beta_D=2\times10^{3}$ (system type I).\\

\begin{figure}
\includegraphics[width=1.0\linewidth]{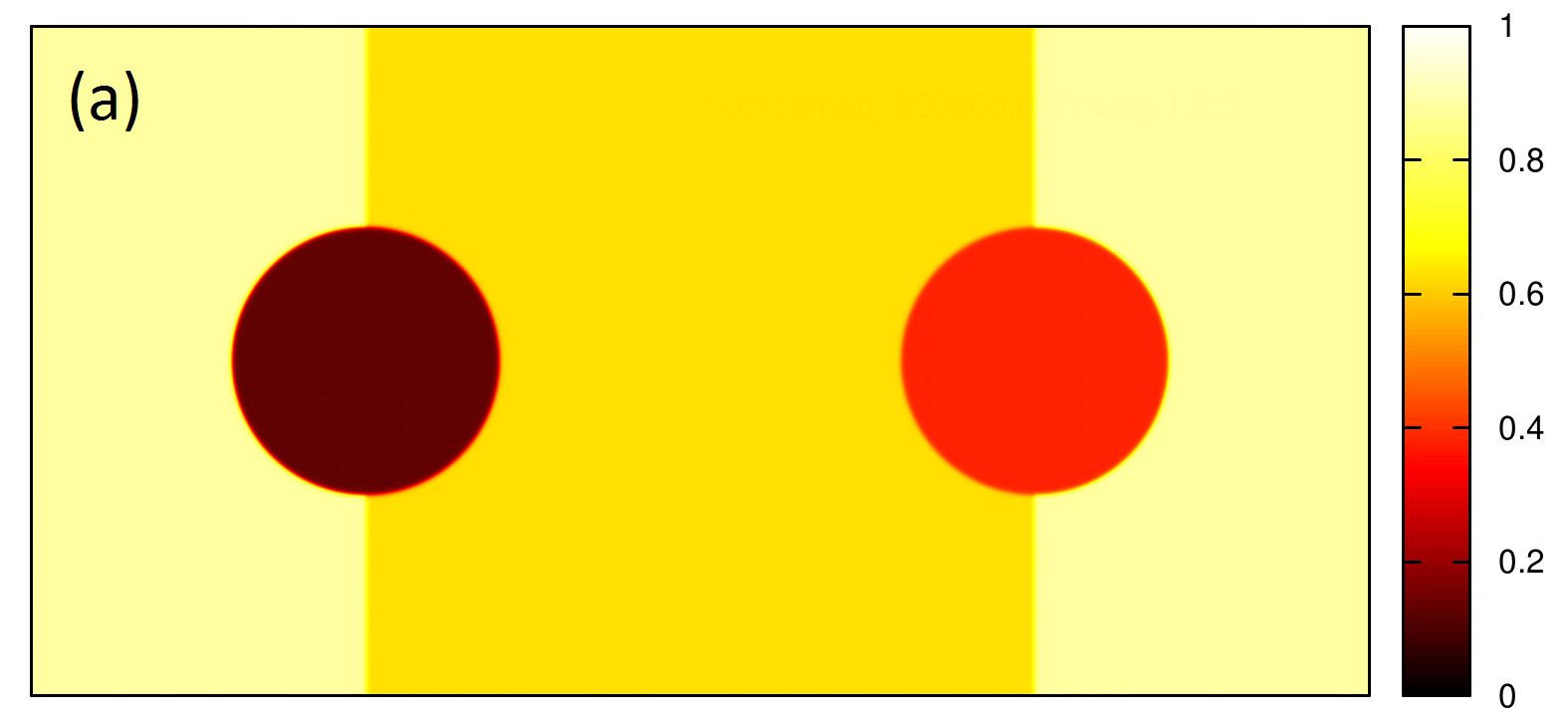}\\
\includegraphics[width=1.0\linewidth]{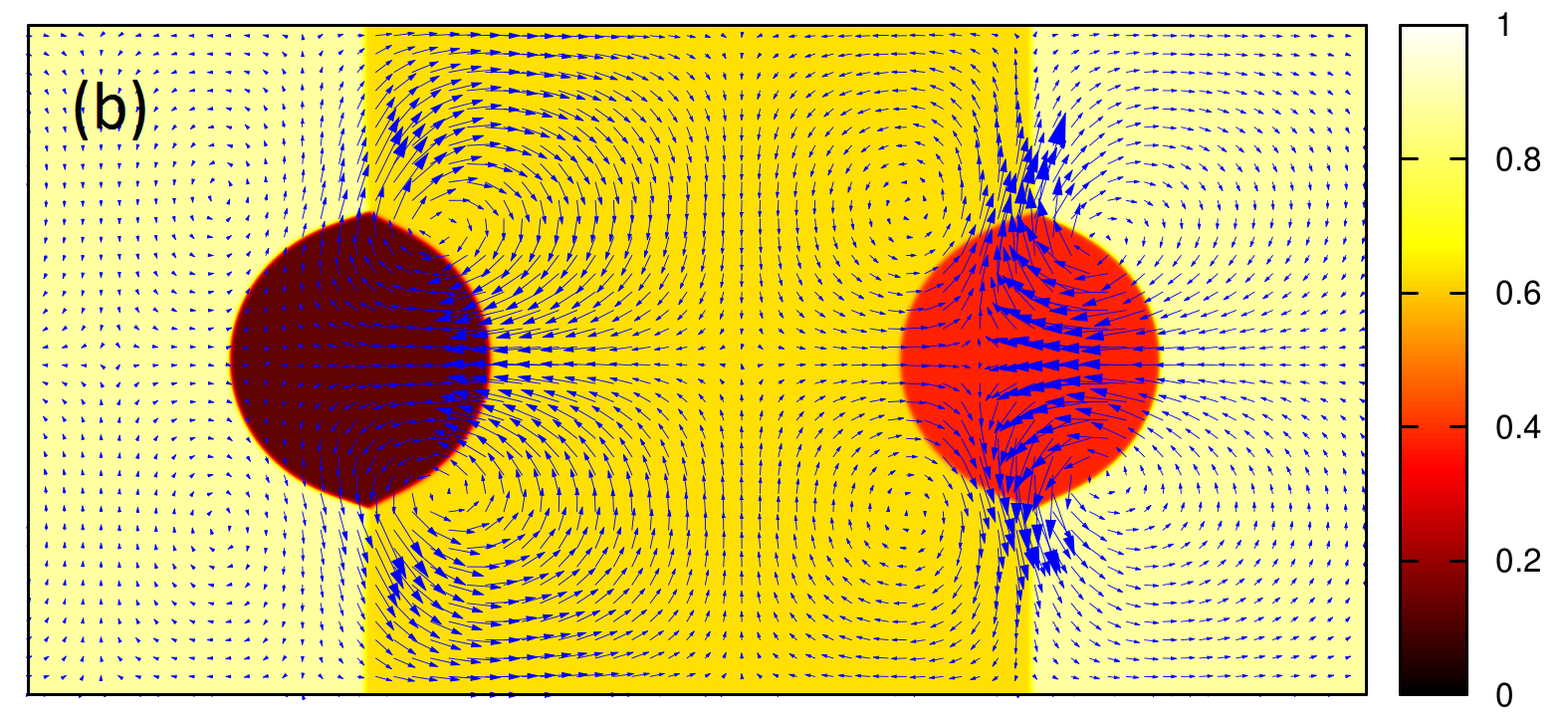}\\
\includegraphics[width=1.0\linewidth]{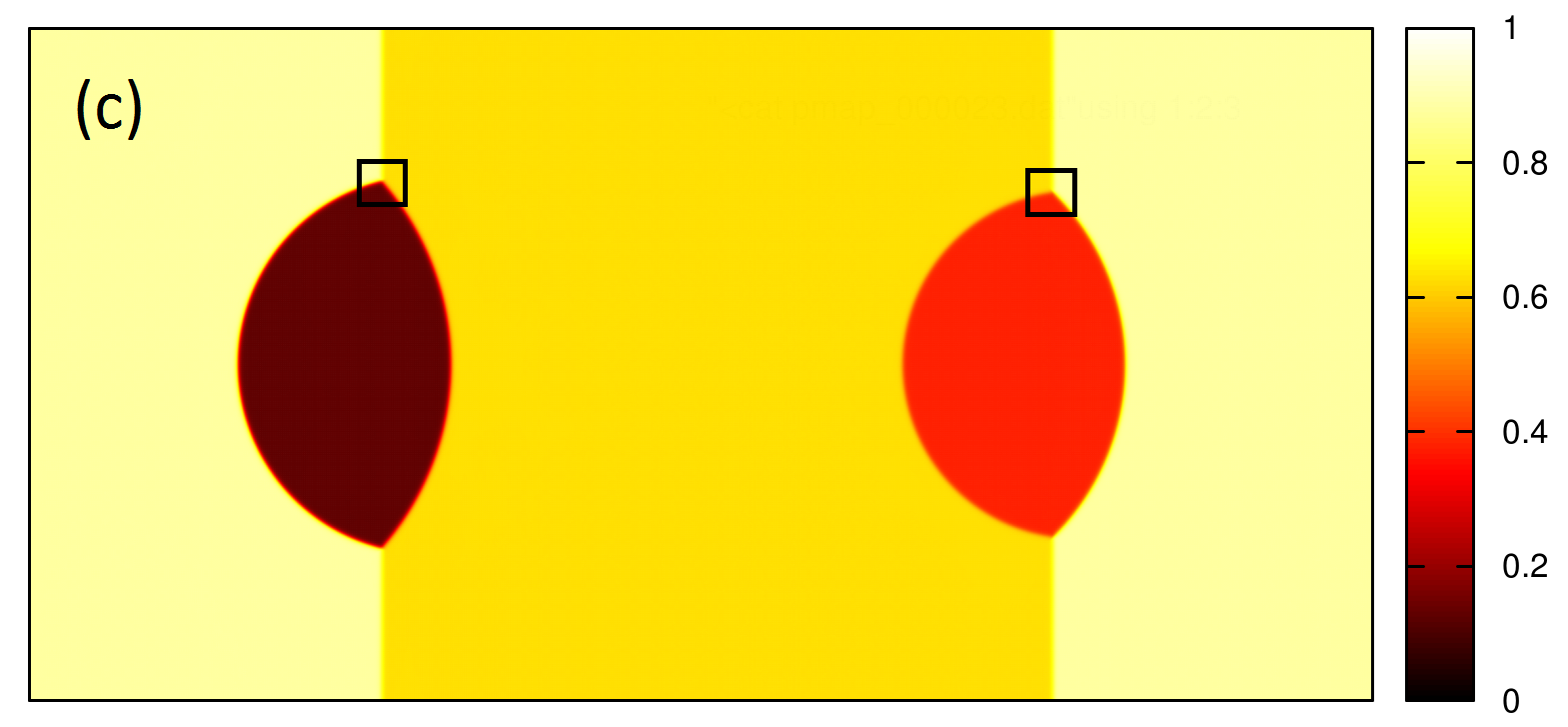}\\
\includegraphics[width=0.49\linewidth]{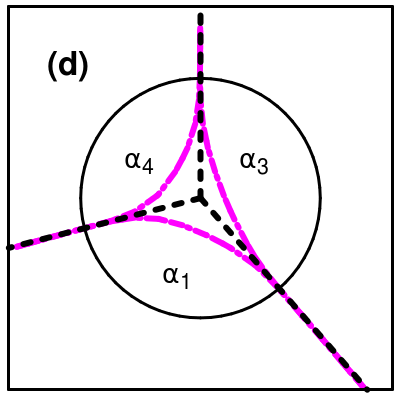}
\includegraphics[width=0.49\linewidth]{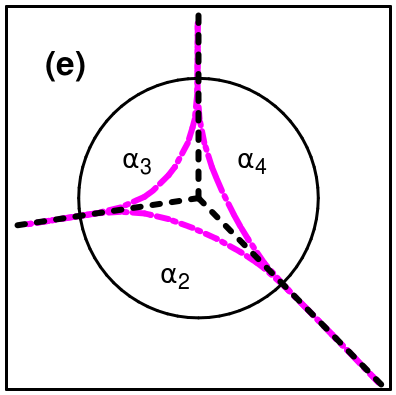}
\caption{Equilibrium contact angles in a 4-component Cahn-Hilliard liquid. The spatial distribution of the scalar variable $h(\hat{\mathbf{r}},\hat{t}):=\sum_{i=1}^4 c_i(\hat{\mathbf{r}},\hat{t})[(i-1/2)/4]$ is displayed at $\hat{t}=0$, $\hat{t}=7.5$, and $\hat{t}=12500$ on panels (a)-(c), respectively. The fluid flow is indicated by vectors on panel (b). Panels (d) and (e) show the $c_i(\hat{\mathbf{r}},\hat{t})=1/2$ contours (solid lines) in the small vicinity of the trijunctions indicated by the small black squares on panel (c), while the dashed lines indicate the corresponding contact angles.}
\end{figure}
The equilibrium contact angles were investigated in a 4-component system first. The dimensionless interfacial tensions were $\hat{\sigma}_{12}=1.0$, $\hat{\sigma}_{13}=1.1$, $\hat{\sigma}_{14}=0.75$, $\hat{\sigma}_{23}=0.9$, $\hat{\sigma}_{24}=1.25$, and $\hat{\sigma}_{34}=1.0$, while the interface thicknesses and the diffusion coefficient were chosen to be {unique}, i.e. $\hat{\delta}_{ij}=\hat{D}_{ij}= 1.0$. $\hat{A}_3=1.0$ was chosen to stabilize the equilibrium binary planar interfaces. The dimensionless viscosities and densities were $\hat{\eta}_1=\hat{\eta}_3=1.0$, $\hat{\eta}_2=0.5$ and $\hat{\eta}_4=2.0$, and $\hat{\rho}_1^0=0.5$, $\hat{\rho}_2^0=1.5$, $\hat{\rho}_3^0=1.25$ and $\hat{\rho}_4^0=0.75$, respectively. {The} initial condition {was} a binary background "sandwich" consisting of two $(3,4)$ planar interfaces, {while} two circular dots of liquids $1$ and $2$ were placed onto these interfaces [see Fig. 1(a)]. In the early stage of the simulation significant fluid flow [the maximal velocity magnitude was $\max(\hat{v})\approx 0.3$] was generated by the non-equilibrium interfaces [see Fig. 1(b)], but after $10^6$ time steps the system was approaching equilibrium [the average velocity magnitude is $\langle\hat{v}\rangle \approx 10^{-4}$, corresponding to 1 pixel shift in $10^6$ time steps]. The equilibrium configuration is shown in Fig. 1(c). The equilibrium contact angles of the trijunctions can be calculated from the condition of mechanical equilibrium by using the interfacial tensions. Accordingly, the equilibrium contact angles for the (1,3,4) trijunction read
\begin{eqnarray}
\alpha_{1} &=& \pi - \cos^{-1} \left( \frac{\hat{\sigma}_{13}^2+\hat{\sigma}_{14}^2-\hat{\sigma}_{34}^2}{2\,\hat{\sigma}_{13}\,\hat{\sigma}_{14}} \right ) \approx 117.92^\circ \\
\alpha_{3} &=& \pi - \cos^{-1} \left( \frac{\hat{\sigma}_{13}^2+\hat{\sigma}_{34}^2-\hat{\sigma}_{14}^2}{2\,\hat{\sigma}_{13}\,\hat{\sigma}_{34}} \right ) \approx 138.49^\circ \\
\alpha_{4} &=& \pi - \cos^{-1} \left( \frac{\hat{\sigma}_{14}^2+\hat{\sigma}_{34}^2-\hat{\sigma}_{13}^2}{2\,\hat{\sigma}_{14}\,\hat{\sigma}_{34}} \right ) \approx 103.59^\circ \enskip .
\end{eqnarray}
Analogously, the contact angles for the $(2,3,4)$ trijunction read $\alpha_{2}\approx 127.59^\circ$, $\alpha_{3}\approx 97.903^\circ$, and $\alpha_{4} \approx 134.51^\circ$. The contact angles can be measured directly from the simulations [see Fig. 1(d) and 1(e)], indicating less than $1.5\%$ relative error compared to the exact values. 
\begin{figure}
\includegraphics[width=0.49\linewidth]{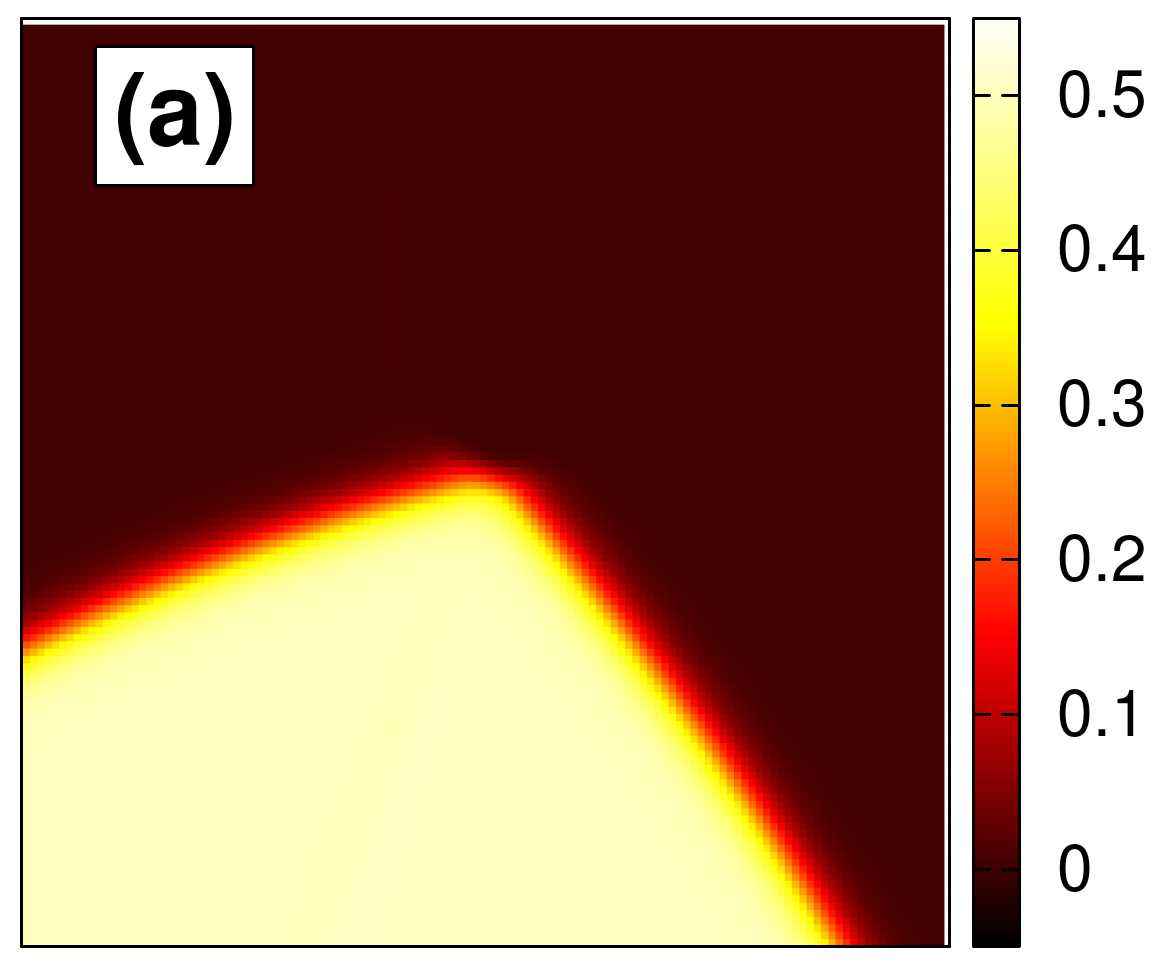}
\includegraphics[width=0.49\linewidth]{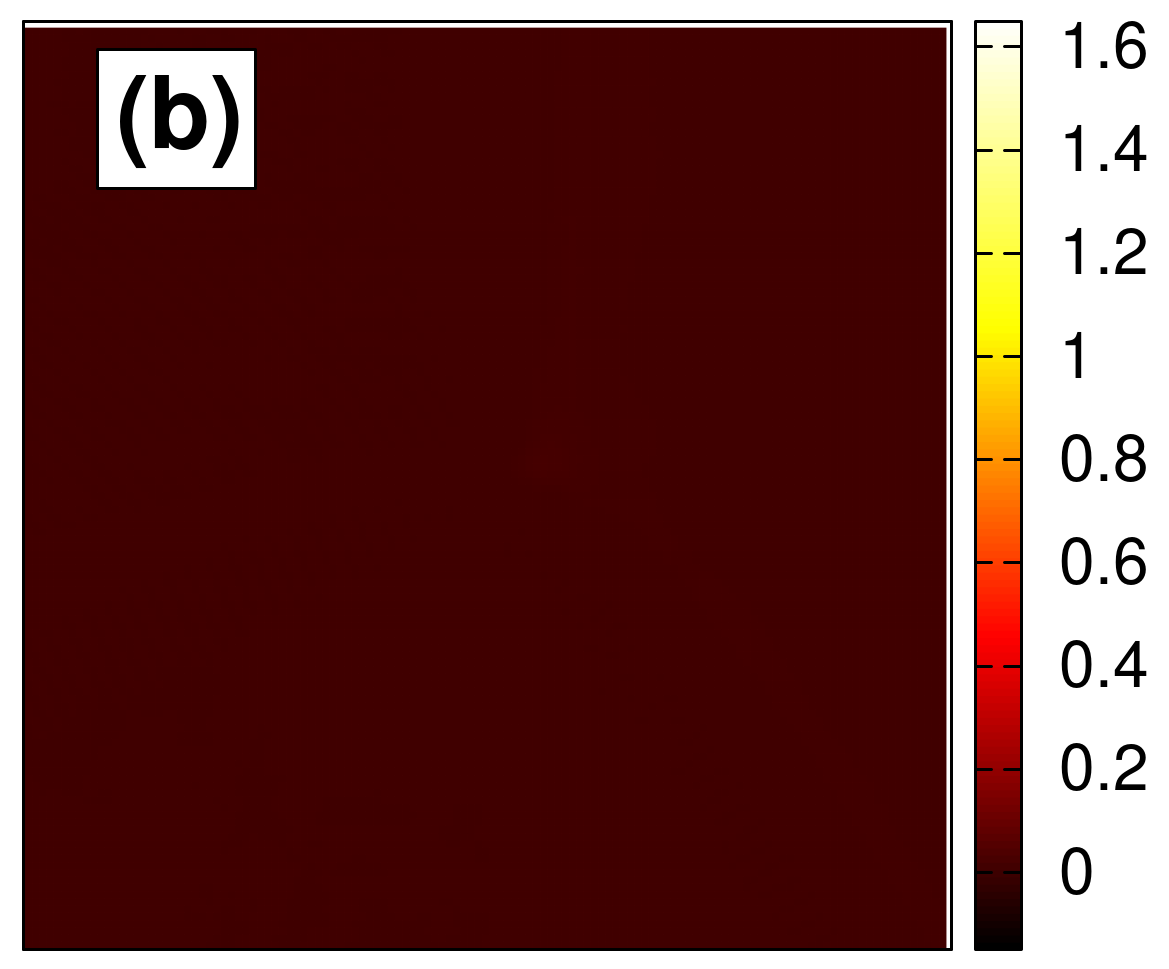}\\
\includegraphics[width=0.49\linewidth]{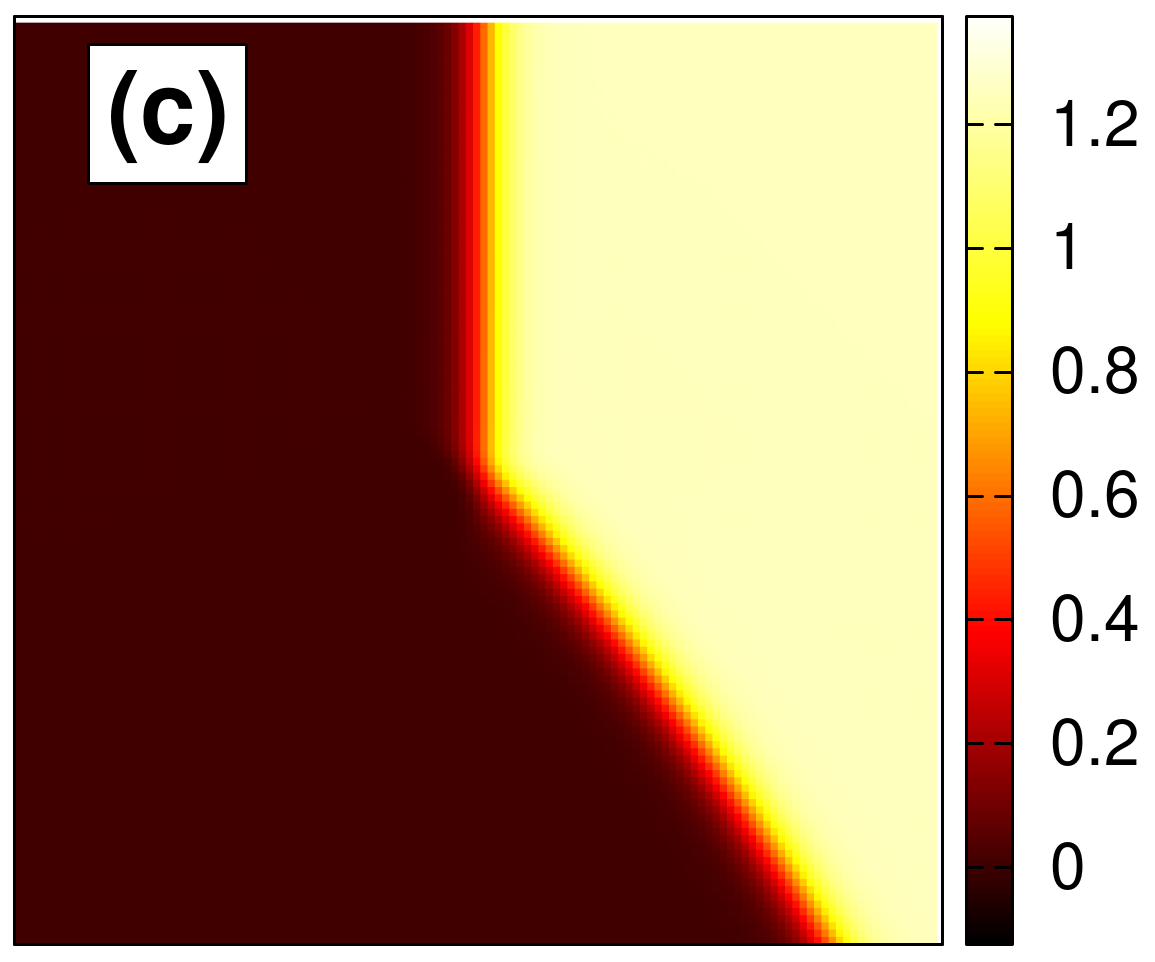}
\includegraphics[width=0.49\linewidth]{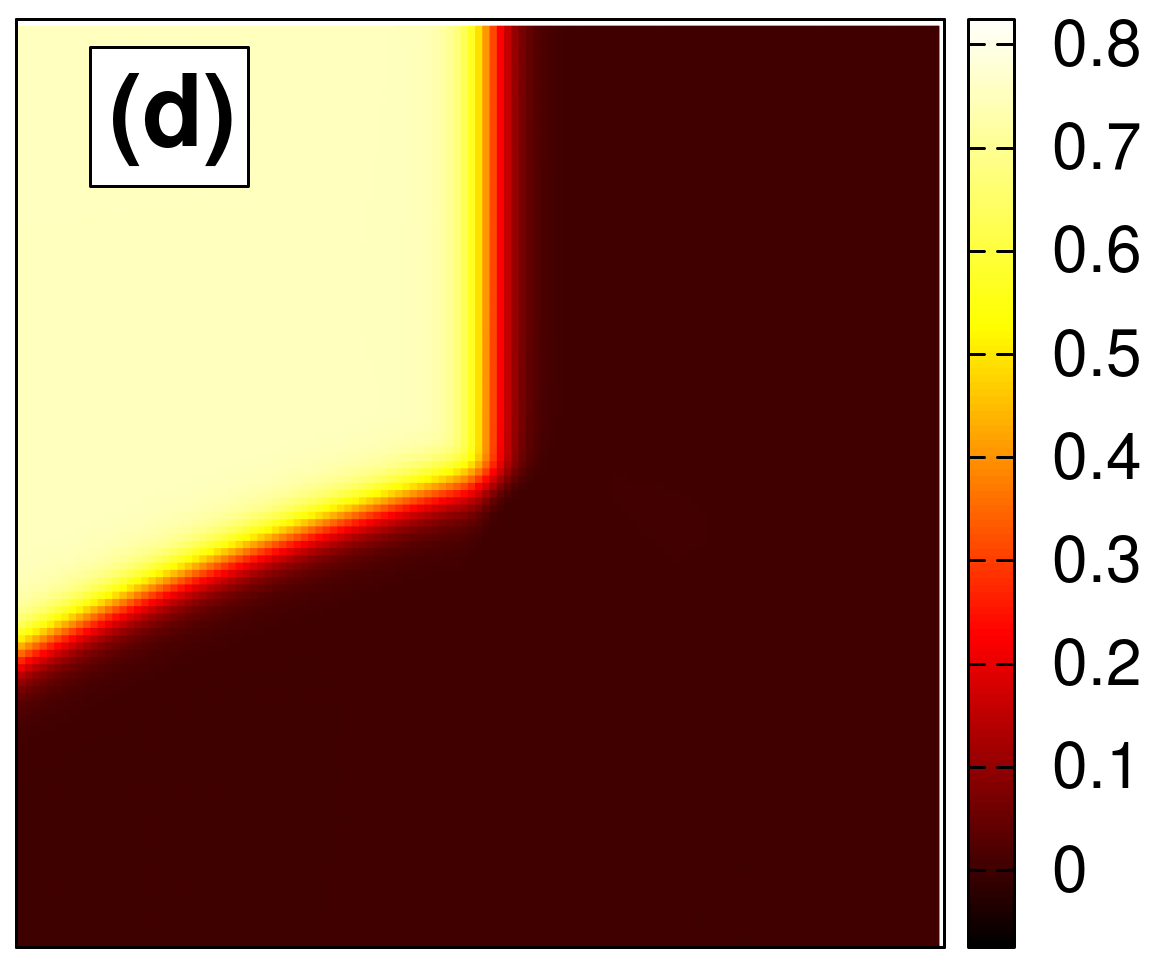}
\caption{Equilibrium trijunction in a 4-component Cahn-Hilliard liquid. Panels (a)-(d) show the individual density fields $\hat{\rho}_i(\hat{\mathbf{r}})$ ($i=1\dots 4$, respectively) at the $(1,3,4)$ trijunction shown in Fig. 1(d). Note the lack of spurious phases.}
\end{figure}
Furthermore, the individual fields (see Fig. 2) show no static spurious phases, i.e. no third component appears at the binary interfaces, even close to the trijunctions. In conclusion, equilibrium is invariant for the density contrast, as expected from Eq. (\ref{eq:condEL}), but we emphasize that this feature follows directly from the special choice of the free energy functional and the mobility matrix: {First,} any solution of the constant density Euler-Lagrange equations, at which the functional derivatives are equal, also satisfies the general Euler-Lagrange equations with an arbitrary constant Lagrange multiplier $\lambda(\mathbf{r})$. Besides, the stationary solutions of the dynamic equations {are} identical to the equilibrium solutions of the free energy functional for the present choice of the mobility matrix. Consequently, the solution of the dynamic equations must converge to an equilibrium solution of the free energy functional. We {also} emphasize that these requirements (for the equilibrium solutions and the mobility matrix) are not fulfilled automatically, and always have to be checked, as pointed out in some recent works \cite{PhysRevB.92.184105,PhysRevE.93.013126}.\\
\begin{figure}
\includegraphics[width=0.49\linewidth]{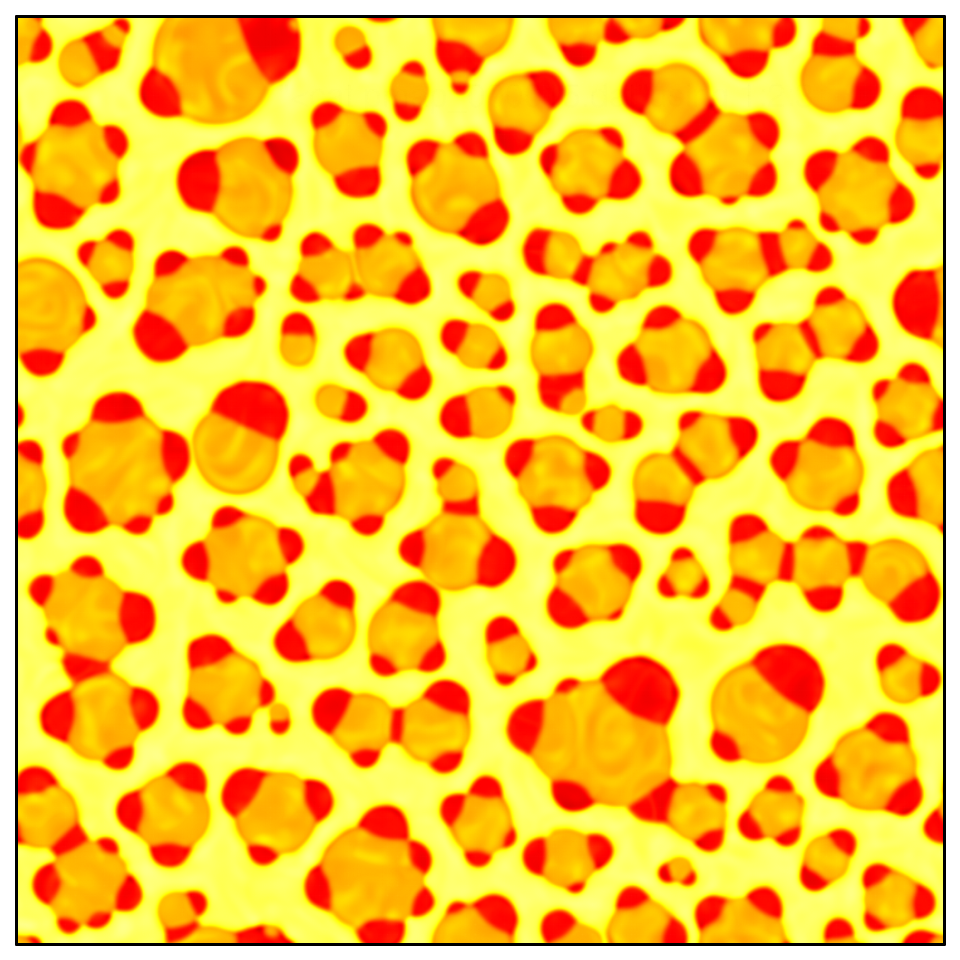}
\includegraphics[width=0.49\linewidth]{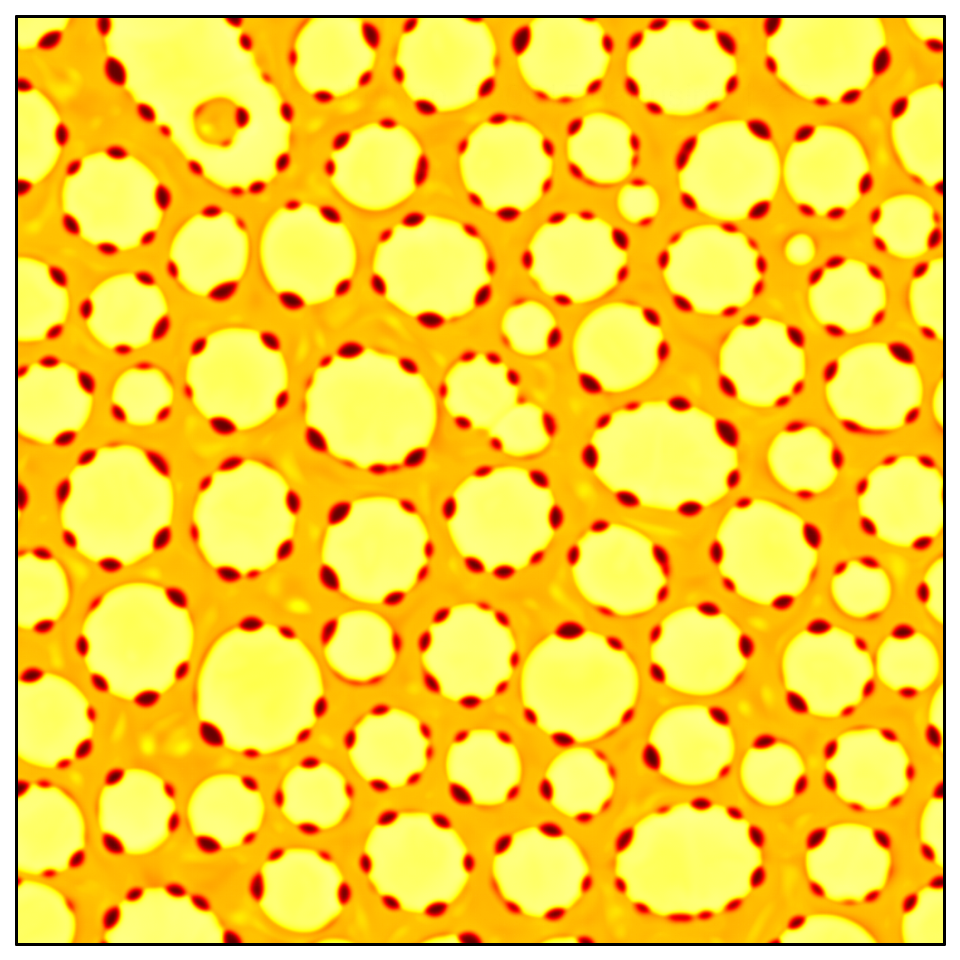}\\
\includegraphics[width=0.49\linewidth]{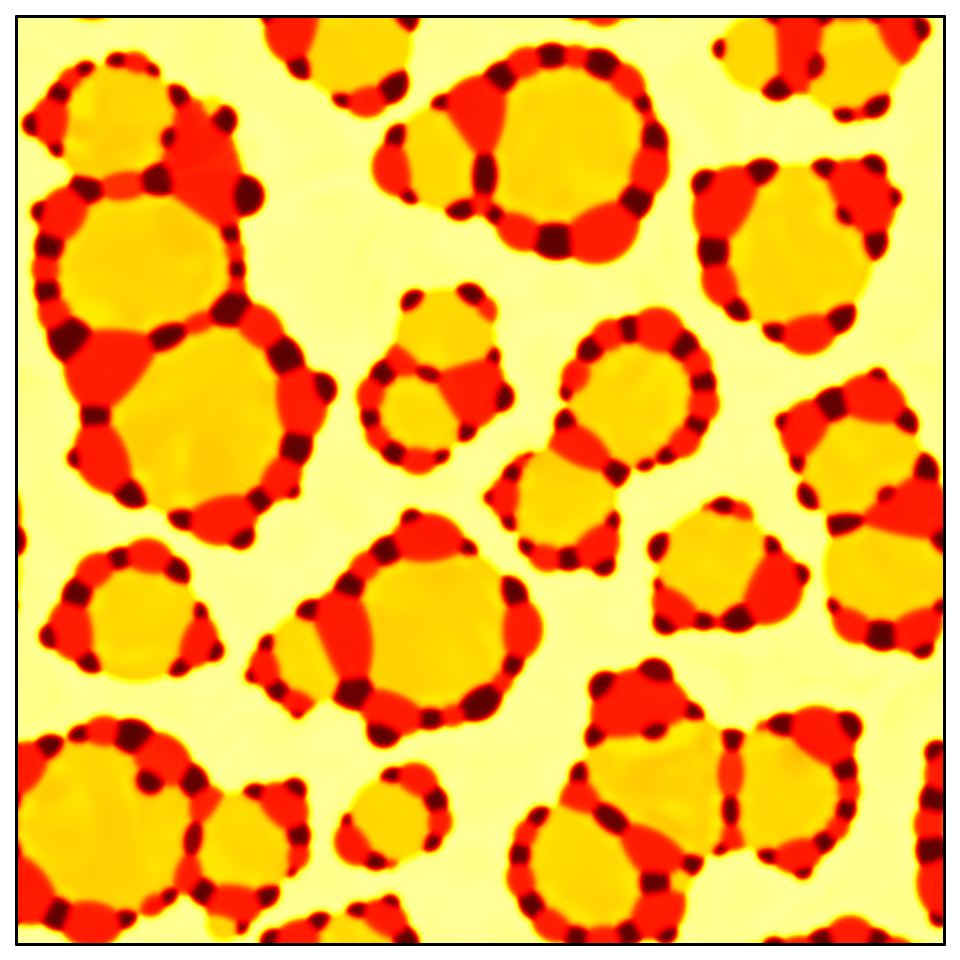}
\includegraphics[width=0.49\linewidth]{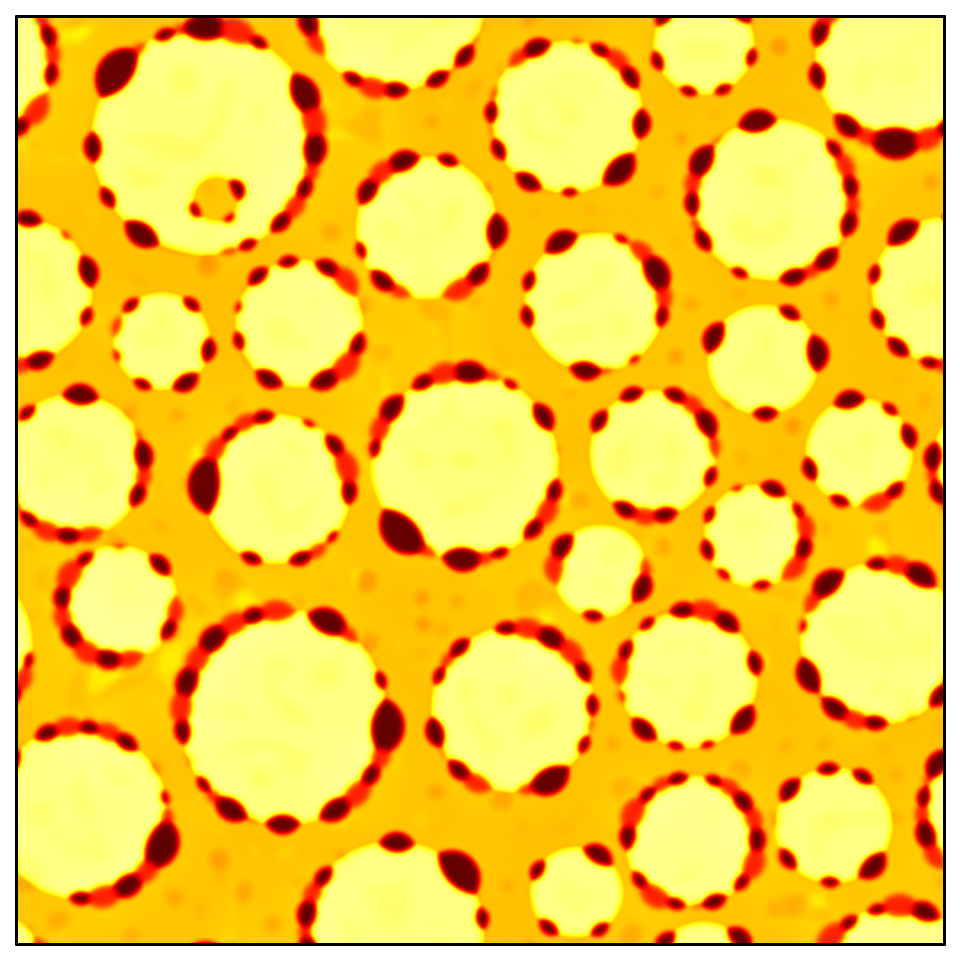}\\
\includegraphics[width=0.49\linewidth]{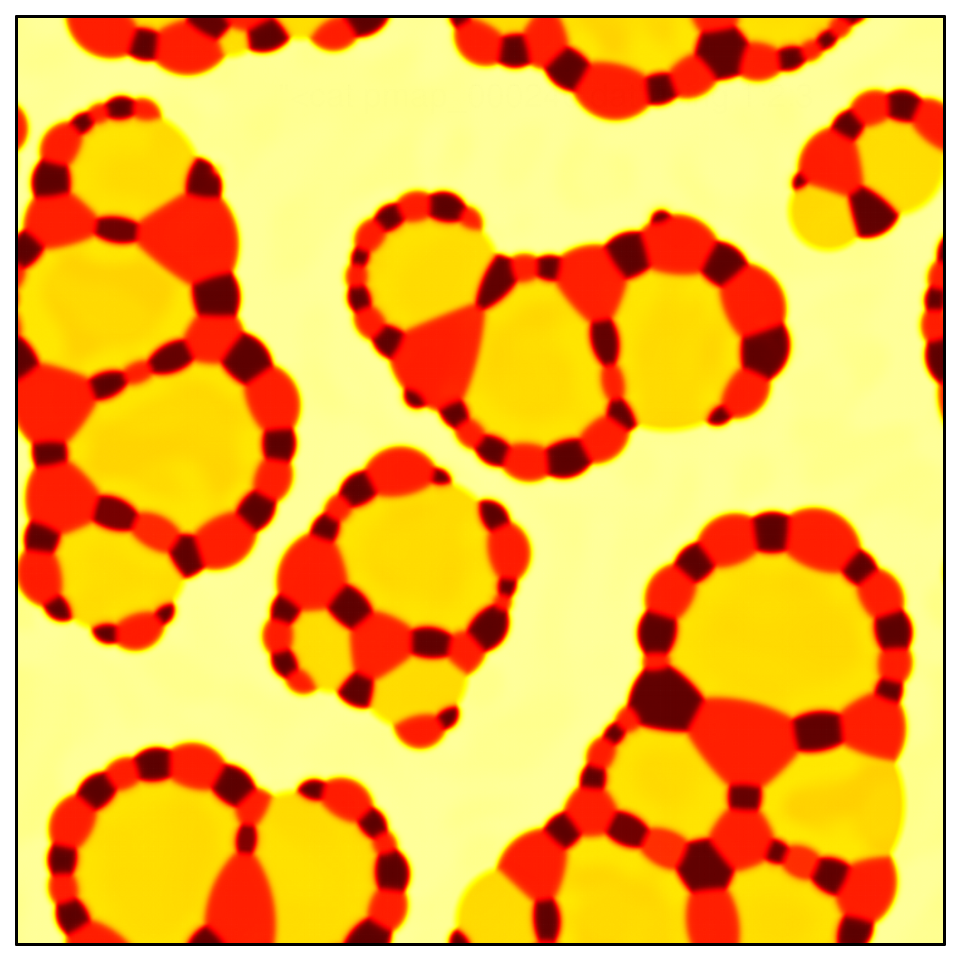}
\includegraphics[width=0.49\linewidth]{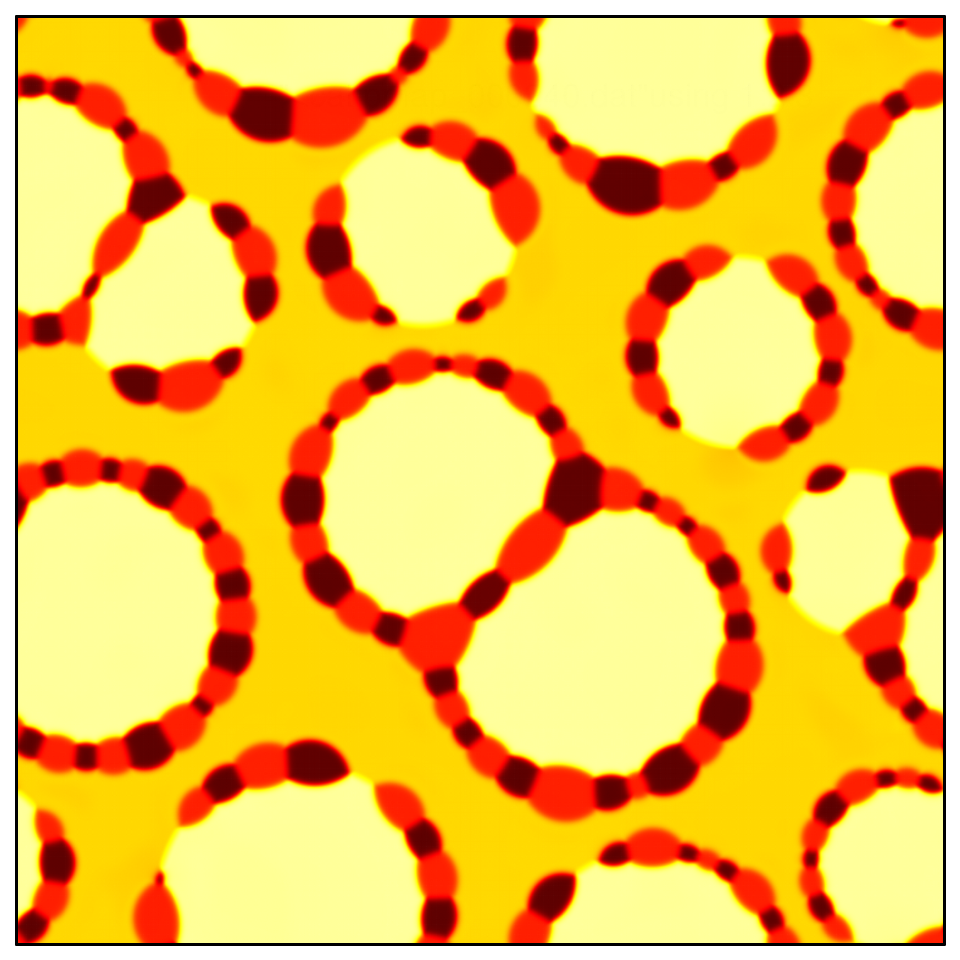}\\
\caption{Phase separation in a 4-component Cahn-Hilliard liquid. Snapshots of simulations taken at time points $\tilde{t}=375$, $\tilde{t}=1500$ and $\tilde{t}=6000$ are shown (from top to bottom, respectively) in case of constant density (on the left) and variable density (on the right). The scalar field defined in the caption of Fig. 1 with the same colorbar is shown.}
\end{figure}

After validating the theory for equilibrium, the effect of density contrast on {phase separation is investigated}. Since pattern formation is a \textit{volumetric} phenomenon, the volume fractions $\phi_i(\mathbf{r},t)\equiv\rho_i(\mathbf{r},t)/\rho_i^0$ {are used} to set up the initial condition. Note that $\phi_i(\mathbf{r},t)$ is also conserved for quasi-incompressible liquids, making thus the constant and variable density simulations comparable. The initial condition reads $\phi_1(\mathbf{r},0)=0.1+a\mathbb{R}[-1,1]$, $\phi_2(\mathbf{r},0)=0.2+a\mathbb{R}[-1,1]$, $\phi_3(\mathbf{r},0)=0.3+a\mathbb{R}[-1,1]$, and $\phi_4(\mathbf{r},0)=1-\sum_{i=1}^3 \phi_i(\mathbf{r},0)$, where $a \ll 1$ and $\mathbb{R}[-1,1]$ is a uniformly distributed random number on $[-1,1]$. Two scenarios were considered, one for constant ($\hat{\rho}_i =1.0$), and another for varying ($\hat{\rho}_1^0=1.0$, $\hat{\rho}_2^0=0.5$, $\hat{\rho}_3^0=1.5$ and $\hat{\rho}_4^0=1.0$) density. All other dimensionless model parameters were chosen to be $1.0$. Snapshots of the simulations are shown in Fig 3. The difference between the constant and variable density case is remarkable: {In case of constant density}, the 3 minority components aggregate in the majority component. The aggregates consist of bubbles of {various sizes} of the dominant minority component connected by {bubble} "chains" of the other two {minority components}. {In case of variable density, however,} the majority component forms individual bubbles  surrounded by chains made of the two minority components. The explanation of the difference relies on the density difference: In {case of constant density}, the effective mobility of the components are equal, i.e. the magnitude of the velocity "reply" to the force density generated by the interfaces is the same in all component. Indeed, comparing the pattern to Fig. 9 of Reference [8] shows qualitative {agreement}. In contrast, a significant density difference between the components has major influence on pattern formation. In {the present} case, {the third} component is the least mobile, since it has the highest density. The formation of the individual islands can be explained by the drop coalescence process. When two bubbles {of component 3} meet, small bubbles (consisting of the other 3 components) stuck {in between} receive high velocity and move fast towards the perimeter of the type 3 bubbles. After the coalescence, the bubbles made of the two minority components have a relatively small chance to meet with each other, therefore, they simply form a chain around the big type-3 bubble, and, consequently, the original majority component forms a background.\\
\begin{figure}
\includegraphics[width=1.0\linewidth]{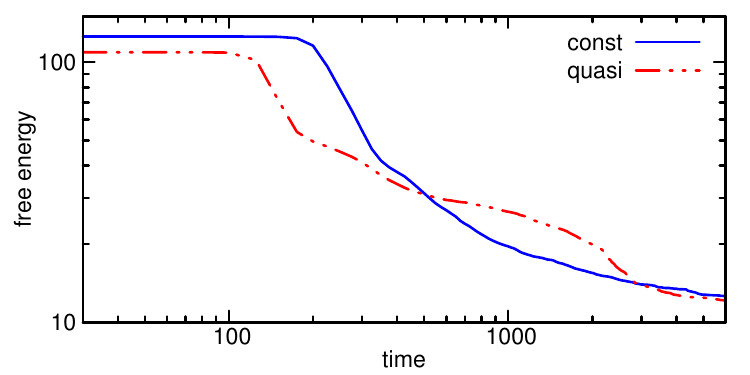}
\caption{Time evolution of the total free energy in phase separating 4-component systems shown by Fig. 3 for constant (solid line) and variable (dash-dotted line) density.}
\end{figure}

Finally the time dependence of the total free energy of the system {is addressed}. According to the second law of thermodynamics, the entropy production rate must be non-negative. This condition is fulfilled for a non-negative viscosity and a positive semi-definite transport matrix appearing in the condition
\begin{equation}
\label{eq:condep}
\sum_{i,j} \nabla\tilde{\lambda}_i L_{ij}(\mathbf{c}) \nabla\tilde{\lambda}_j \geq 0 \enskip .
\end{equation}   
Note that Eq. (\ref{eq:condep}) prescribes the positive semi-definiteness of $\mathbb{L}$, independently from the form of the forces, which is just a consequence of the linearity postulate. In addition, the contribution emerging from the Navier-Stokes equation is solely attributed to the viscosity. {Following} these arguments, a pressure stable numerical scheme {is} necessarily energy minimizing even in the general incompressible case. It is demonstrated in Fig. 4, showing  monotonously decreasing total free energy during the phase separation process shown in Fig. 3. Summarizing, we have a numerical evidence that both constant density and quasi-incompressible systems described by the  dynamic equations and the free energy construction presented in this work obey the second law of thermodynamics.

\section{Summary}

Herein we derived dynamic equations describing the time evolution of an isothermal multicomponent liquid mixture, {where} the free energy is given in the framework of {gradient theories}. The model integrates the fundamental equations of fluid mechanics and the principles of irreversible thermodynamics, and can be straightforwardly applied for both compressible and general incompressible systems. Since the mathematical formulation is compact, {the model can be applied for many liquid systems of practical importance featuring variable density}. It is important to mention, that the extension of the model with phase transformation and variable temperature {is also possible} \cite{PhysRevE.64.051601,PhysRevE.67.026117}. The only difficulty in utilizing the model for non-trivial incompressible systems [i.e. when $\varrho(\mathbf{c})\neq const$] is the implementation of the pressure solver, which, however, has been solved at least for quasi-incompressible systems. As it has been demonstrated, the solver is pressure stable / energy minimizing, without using complex discretization schemes. It has also been demonstrated, that the choice of the free energy functional and the mobility matrix must be compatible with the dynamic equations, in order to preserve the structure of constant density equilibrium solutions in case of variable density, allowing thus the analytic calibration of the model. In addition, we observed that the effect of density contrast on pattern formation could be crucial even in case of a density contrast of 2.

\section*{Acknowledgments}

The author wish to thank Bj\o rn Kvamme and Juri Selv\aa g, Department of Physics and Technology, Universiy of Bergen for the valuable discussions. The work has been supported by the VISTA basic research programme Project
No. 6359 “Surfactants for water-CO2-hydrocarbon emulsions
for combined CO2 storage and utilization” of the Norwegian
Academy of Science and Letters and the Statoil.

\bibliography{./fluids}

\end{document}